\documentclass[sigconf]{acmart}

\usepackage{booktabs} 

\setcopyright{rightsretained}
\usepackage{graphicx}
\usepackage{amsmath}
\usepackage{amssymb}
\usepackage{bm}
\usepackage{booktabs}
\usepackage{multirow}

\copyrightyear{2017}
\acmYear{2017}
\setcopyright{acmcopyright}
\acmConference{ThematicWorkshops'17}{}{October 23--27, 2017, Mountain View, CA, USA}
\acmPrice{15.00}
\acmDOI{http://dx.doi.org/10.1145/3126686.3126720}
\acmISBN{ISBN 978-1-4503-5416-5/17/10}

\begin{document}
\title{Learning Social Image Embedding with Deep Multimodal Attention Networks}

\author{Feiran Huang}
\orcid{1234-5678-9012}
\affiliation{%
  \institution{Beihang University}
  \city{Beijing}
  \state{China}
}
\email{huangfr@buaa.edu.cn}

\author{Xiaoming Zhang}
\affiliation{%
  \institution{Beihang University}
  \city{Beijing}
  \state{China}
}
\email{yolixs@buaa.edu.cn}

\author{Zhoujun Li}
\affiliation{%
  \institution{Beihang University}
  \city{Beijing}
  \state{China}
}
\email{lizj@buaa.edu.cn}

\author{Tao Mei}
\affiliation{%
  \institution{Microsoft Research}
  \city{Beijing}
  \state{China}
}
\email{tmei@microsoft.com}

\author{Yueying He}
\affiliation{%
  \institution{National Computer Network Emergency Response Technical Team/Coordination Center of China}
  \city{Beijing}
  \state{China}
}
\email{hyy@cert.org.cn}

\author{Zhonghua Zhao}
\affiliation{%
  \institution{National Computer Network Emergency Response Technical Team/Coordination Center of China}
  \city{Beijing}
  \state{China}
}
\email{zhaozh@cert.org.cn}


\begin{abstract}

Learning social media data embedding by deep models has attracted extensive research interest as well as boomed a lot of applications, such as link prediction, classification, and cross-modal search. However, for social images which contain both link information and multimodal contents (e.g., text description, and visual content), simply employing the embedding learnt from network structure or data content results in sub-optimal social image representation. In this paper, we propose a novel social image embedding approach called Deep Multimodal Attention Networks (DMAN), which employs a deep model to jointly embed multimodal contents and link information. Specifically, to effectively capture the correlations between multimodal contents, we propose a multimodal attention network to encode the fine-granularity relation between image regions and textual words. To leverage the network structure for embedding learning,  a novel Siamese-Triplet neural network is proposed to model the links among images. With the joint deep model, the learnt embedding can capture both the multimodal contents and the nonlinear network information. Extensive experiments are conducted to investigate the effectiveness of our approach in the applications of multi-label classification and cross-modal search. Compared to state-of-the-art image embeddings, our proposed DMAN achieves significant improvement in the tasks of multi-label classification and cross-modal search.

\end{abstract}




\maketitle


\section{INTRODUCTION}

With the rise of social network, the data like social images containing both content and link information has becoming more and more popular in miscellaneous social media (e.g., Facebook, Flickr, and Twitter), which requires an effective method to process and analyse them. Learning an efficient representation to capture the content and link information facilitates a good solution to it. The learnt social media data representations have gained great success in content-centric and network oriented applications, such as multilabel image classification, cross-modal image search, and link prediction. Therefore, how to represent the data to a vectorized space, also called social media data embedding, has been increasingly attracting attention in both academia and industry, which subsequently posits a significant challenge to social image embedding: can both the multimodal contents and links be combined for embedding learning?

\begin{figure}[t!]
  \centering
  \begin{minipage}[b]{0.49\textwidth}
  \includegraphics[width=0.99\textwidth]{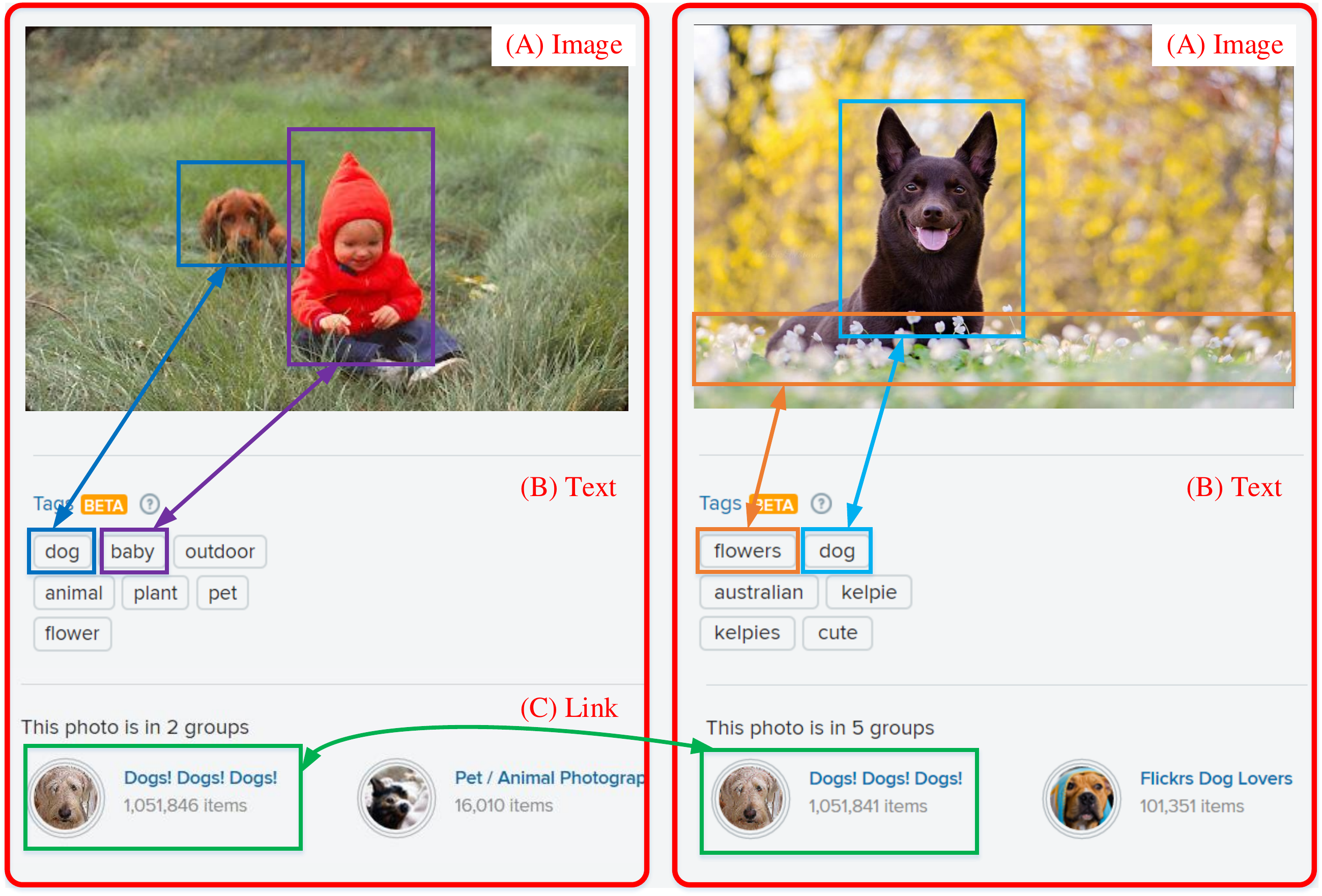}
  \end{minipage}
  \caption{An example of social image: (A) a shared photo, (B) tags provided by the owner, (C) albums and galleries that contain many similar images; (A) and (B) constitute the multi-modal content of social image and (C) forms the link between images.}
  \label{fig:flickr_example}
\end{figure}

It is of great challenge to deal with the social media data for embedding. First, the social images contain diverse patterns of manifestations such as image and text description. These data modalities are heterogeneous in the feature spaces. Second, there exists link relation among the data, which indicates that an efficient embedding should leverage both the nonlinear network information and data content to learn a unified representation. Third, the amount of social images on social networks has increased exponentially. Therefore, it needs an efficient method to effectively learn the embedding from the large amount of data. Figure \ref{fig:flickr_example} gives an example of social image.

Most of the existing social media data embedding methods can be categorized into two classes, i.e., network-based and content-based. The network-based embedding methods learn a representation for the nodes to capture the netweork structure, which includes the shallow model based methods, e.g., GraRep \cite{2015Grarep-Cao} , Line \cite{2015LINE-Tang} and PPNE \cite{2017PPNE-Li}, and the deep model based methods, e.g., SDNE \cite{2016Structural-Wang}. These methods mainly use the proximity information in the network to learn the embedding, which ignores the content contained by each node. The content based methods mainly use a supervised or semi-supervised method to learn a joint representation for image and text \cite{2016Joint-Ren, 2013Zero-Norouzi}, which ignores the linkage among data and the fine-granularity relation between different data modalities. Though HNE \cite{2015Heterogeneous-Chang} is proposed to combine both the network and content for embedding learning, it models different data modalities independently and the learning process is time-consuming.

Meanwhile, the multi-modal, heterogeneous and interconnected characteristics of social media data provide clues for social image embedding. First, social images does not exist in isolation, but are linked explicitly or implicitly. Though the interconnection violates the independently and identically distributed assumption in most statistical machine learning algorithms, both content and interconnection information can be exploited to complement each other for better solutions. Second, though different modalities of content are heterogeneous, there exists fine-granularity relation between them. For example, as shown in Figure \ref{fig:flickr_example}, some words, such as ``dog'', ``baby'', and ``flowers'', are related with the specific regions in the corresponding images. If the relation is parsed accurately, these words and visual regions can be modelled jointly in an intimate fashion and the salient features are allowed to come to the forefront as needed.

\begin{figure*}[tb]
  \centering
  \begin{minipage}[b]{0.98\textwidth}
  \includegraphics[width=0.99\textwidth]{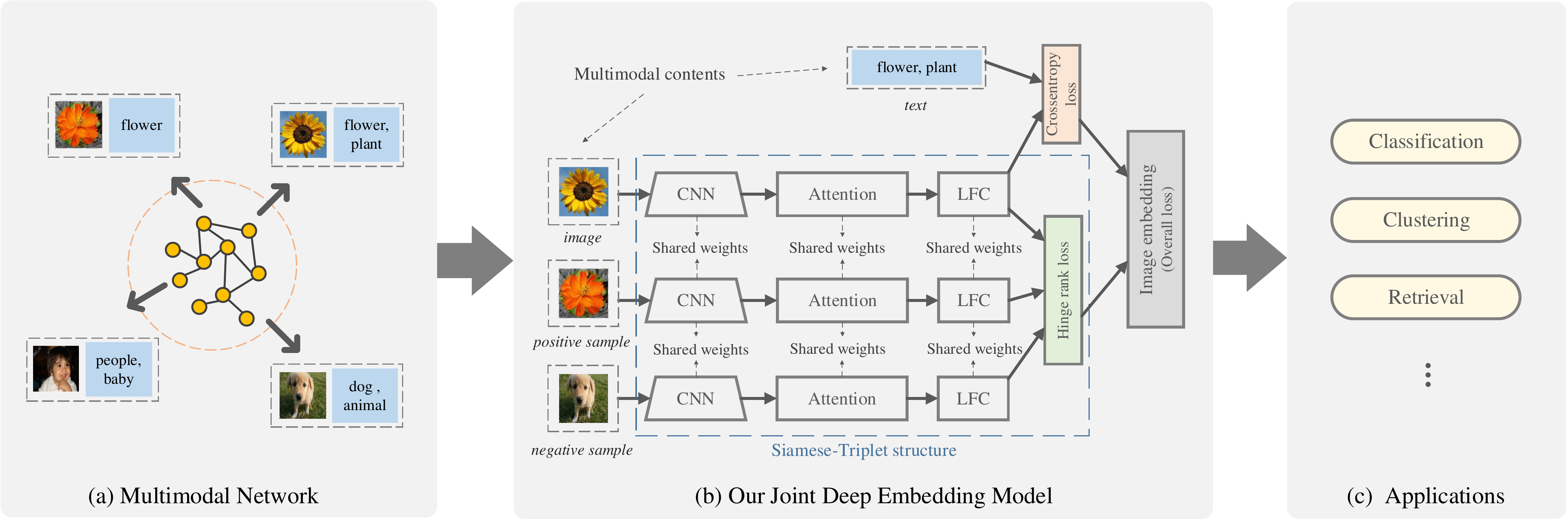}
  \end{minipage}
  \caption{The framework of DMAN for social image embedding: (a) Multimodal network where each node contains two modalities of image and text. Each node is linked with others through some kind of relationship; (b) Our joint deep model to establish a joint representation of the multimodal network. The whole network is based on Siamese-Triplet structure which is composed of three identical subnetworks that share the same parameters. Specifically, in each subnetwork, we use CNN layers to extract image features and employ Attention layer and Locally Fully Connected (LFC) layers for visual-textual alignments. Hinge rank loss and Crossentropy loss are used to learn network information and multimodal contents respectively; (c) Various applications can be conducted on the learnt embedding.}
  \label{fig:framwork}
\end{figure*}

To tackle the above challenges, we propose to take advantage of the link information and multimodal contents in social images for embedding. In particular, we investigate: (1) how to capture the fine-granularity relation between different data modalities in the learnt representation; (2) how to combine the link information for social image embedding. Our solutions to these questions result in a novel approach called Deep Multimodal Attention Networks (DMAN) for social image embedding. It aims at learning social image embedding that can encode both the multimodal contents and network structure based on a joint deep model. The framework is illustrated in Figure \ref{fig:framwork}. A visual-text attention model is proposed to explore the fine-granularity relation between different data modalities for social image embedding, in which the alignment between image regions and textual words is leveraged to prevent the model from being dominated by single modality. To combine the link information for embedding, the Siamese-Triplet neural network architecture built on Convolution Nerural Network (CNN) is proposed to model the network structure. Then, a joint model is proposed to integrate the two components and embed the multimodal contents and link information into a unified vector space. To improve the efficiency of model inference, we apply the positive and negative sampling method on the Triplet network, which substantially reduce the time complexity of the optimization solution. The main contributions are summarized as follows:
\begin{itemize}
\item Different from traditional data embedding methods, we investigate the problem of learning linkage-embedded social image embedding, where the learned embedding can well capture both the multimodal contents and network structure. Our approach is unsupervised and task independent, which makes it suitable for many network orientated and multi-modal data based data mining applications.
\item We propose a joint deep model (DMAN) to address the challenges of combing content and links for embedding learning, where two models are proposed to capture multimodal contents and network structure respectively, with a deep model to integrate them.

\item We conduct extensive experiments to compare the proposed model with several state-of-the-art baselines on 3 real-world datasets. The experimental results demonstrate the superiority of the proposed model.

\end{itemize}

The rest of this paper is organized as follows. Section 2 summarizes the related work. Section 3 introduces our social image embedding model in details. Section 4 presents the experimental results. Finally we conclude in Section 5.


\section{RELATED WORK}

Learning a representation of social media data has attracted great research attention recently. Most of existing methods can be categorized into two classes, i.e., network-based and content-based.

The network-based methods embed network into a low dimensional space, i.e. learn a vector representation for each node, with the goal of reconstructing the network in the learned embedding space. Most of the network embedding methods adopt the shallow model, e.g., DeepWalk \cite{2014Deepwalk-Perozzi}, and GraRep \cite{2015Grarep-Cao}, etc. DeepWalk \cite{2014Deepwalk-Perozzi} learns the latent representations of the nodes of a social network from truncated random walks in the network. GraRep \cite{2015Grarep-Cao} further extends DeepWalk to utilize high-order proximities. These methods typically first construct the affinity graph based on the feature vectors and then solve the leading eigenvectors of the graph matrices to infer the node embedding. Despite these approaches have achieved certain performance for network embedding, they all adopt the shallow model which is difficult to effectively capture the non-linear structure of the underlying network. Recently, some deep model based methods have been proposed for network embedding. SDNE \cite{2016Structural-Wang} is a semi-supervised deep model that exploits the first-order proximity and second-order proximity to characterize the local and global network structure. However, these network based models only consider the link information and cannot be directly applied to the social media data scene where nodes have abundant content and properties. To combine the content for embedding, HNE \cite{2015Heterogeneous-Chang} is proposed to embed heterogeneous network using a deep model, in which each node is an image or a text document. However, in social media data, each data object usually contains multi-modal contents, and the internal relation between different data modalities makes it not suitable to consider them independently.

The content-based methods learn a joint representation by exploring the correlation between the multimodal contents for specific applications. Some of these methods leverage semantic information from unannotated text data to learn semantic relationships between labels, and explicitly map images into a rich semantic embedding space \cite{2013DeViSE-Frome, 2013Zero-Norouzi}. \cite{2016Joint-Ren} proposes a Gaussian visual-semantic embedding model for joint image-text representation, which leverages the visual information to model text concepts as Gaussian distributions in semantic space. To support user-centric applications, such as image recommendation and reranking, \cite{2015Learning-Liu} proposes to learn image representation to capture both semantic labels and user intention labels. Meanwhile, there are also many works using the multi-modal learning method to learn the correlation between images and text to support various applications, such as image caption \cite{2014Deep-Mao, 2015Deep-Karpathy, 2015Show-Xu} and visual question answering \cite{2015Exploring-Ren, 2016Stacked-Yang}. Canonical Correlation Analysis (CCA) \cite{2004Canonical-Hardoon} and its kernel version (KCCA) \cite{2004Canonical-Hardoon} are the popular methods used in many works \cite{2014Multi-Gong, 2014Improving-Gong, 2014Fisher-Klein}, which finds a projection to maximize the correlation between the vectors projected from different views. Deep canonical correlation analysis (DCCA) \cite{2013Deep-Andrew, 2015Deep-Yan} is a deep model based CCA, which is applicable to high dimensional representation of image and text. However, these content based methods mainly learn the embedding to capture the correlation between different modalities of content, which cannot effectively explore the links among data to improve the embedding learning.

\section{Learning Social Image Embedding}

\subsection{Problem Statement}

Before the problem formulation, we define the notations used in the paper. In this paper, a set of social images is defined as a multi-modal network with each node containing multimodal contents and one or multiple types of links. As a mathematical abstraction, we define an undirected graph $\bm{G} = (\bm{\mathcal{V}},\bm{\mathcal{E}})$, where $\bm{\mathcal{V}}=\{{\mathcal{V}_1},...,{\mathcal{V}_N}\}$ is a set of nodes and $\bm{\mathcal{E}}$ is a set of edges. An edge $\mathcal{E}_{ij},\forall i,j \in \{1,2,...,N\}$ belongs to the set $\bm{\mathcal{E}}$ if and only if an undirected link exists between $\mathcal{V}_i$ and $\mathcal{V}_j$. For further simplification to comprehend, we assume two types of objects in the network: image($V$) and text($T$) and each node contains a pair of these two types of objects. Then $\forall i,\mathcal{V}_i = \{V_i,T_i\}$, where an image is presented as a squared tensor format as
$V_i\in \mathbb{R}^{c \times h \times w}$ ($c$, $h$, and $w$ denote the number of channels, the height, and the width of the image), and the text content is represented as $T_i\in \mathbb{R}^{L}$ ($L$ denotes the size of the tag vocabulary).

Figure \ref{fig:framwork} illustrates the framework of our approach. In detail, to encode the linkage between social images, the Siamese-Triplet neural network is proposed to model the relationship among a triplet of images, i.e., a given image, a positive image which is a randomly sampled image linked to it, and a negative image which is a randomly sampled image with no link to it. The Siamese-Triplet neural network consists of three identical base networks which share the same parameters, with a hinge rank loss to learn the rank of the positive and negative images. To capture the fine-granularity relation between image regions and textual words, we propose a multimodal attention networks model, which assigns reasonable attention weights between the input words and the visual regions for a given social image. To combine the content and network for embedding, then a joint deep model is proposed to integrate the two components by simultaneously optimizing them. Since the number of links in the network is exponential to the number of nodes, directly optimizing the object function by updating the whole network in each iteration will leads to an exponential complexity. Therefore, we propose a positive and negative image sampling method to decrease the complexity of training, which randomly sample a positive and $K$ negative images for each image in the inference process, resulting in a linear complexity .

\subsection{Siamese-Triplet Neural Network Model} \label{sub:siamese}

\begin{figure}[htb]
  \centering
  \begin{minipage}[b]{0.4\textwidth}
  \includegraphics[width=0.95\textwidth]{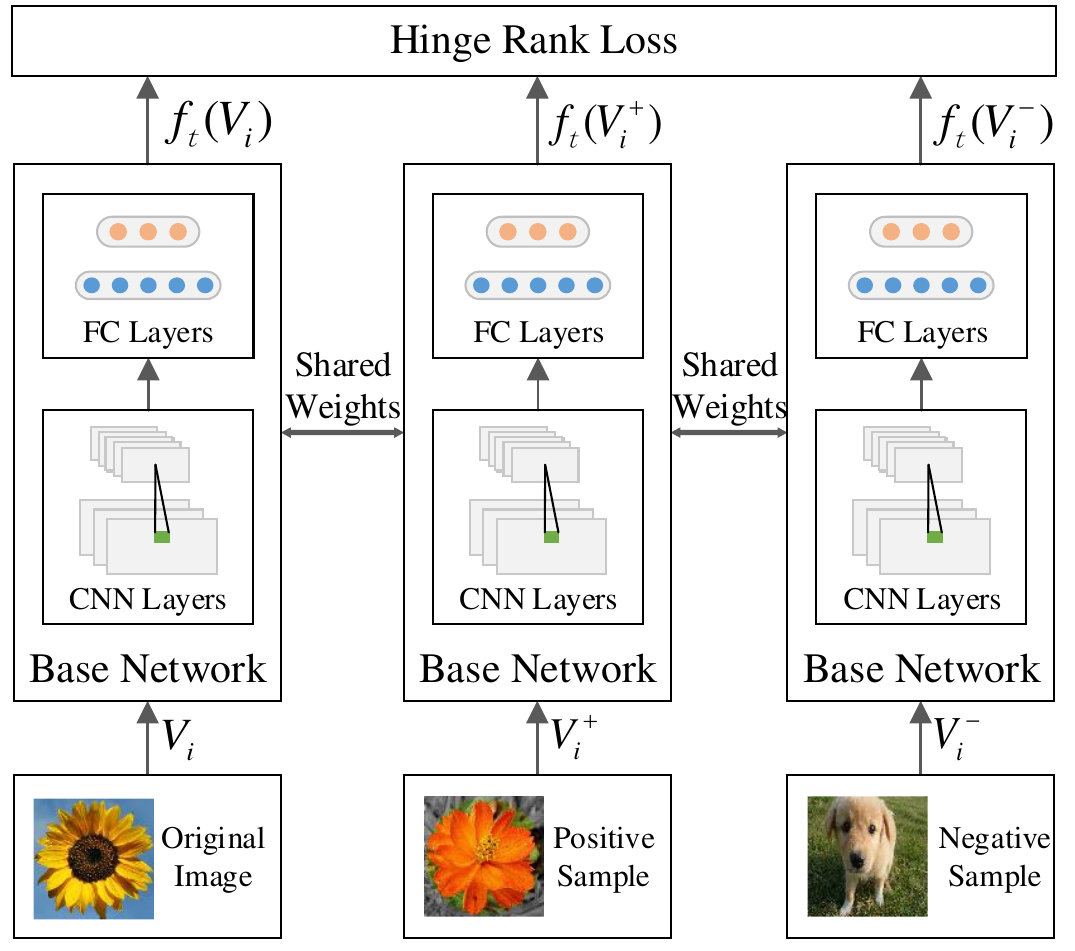}
  \end{minipage}
  \caption{An illustration of the Siamese-Triplet network structure. The base networks consist of CNN and FC layers.}
  \label{fig:tripnet_network}
\end{figure}

Siamese-Triplet architecture is effective to model the network structure, which contains three identical subnetworks sharing the same configuration. Therefore, fewer parameters and less data are required to train it \cite{2015Learning-Wohlhart, 2015Unsupervised-Wang, 2016Learning-VijayKumarB.}. To capture the nonlinear structure of network, we propose a Siamese-Triplet neural network based on deep model for social image embedding as shown in Figure \ref{fig:tripnet_network}. First, we build a deep Convolution Neural Network (CNN) with an addition of several Full Connected (FC) layers as our base network to learn the features of every image. To encode the network information, we then build a network with Siamese-Triplet architecture over the base networks. Usually, a node is more similar to the linked nodes than to a random node. We use the Siamese-Triplet structure to capture the ranking information of the three nodes. Therefore, for a given image, we sample a positive image which has a link to it and a negative image which has no link to it. The three images compose the inputs of the Triplet network.

We use $f_{t}(\cdot)$ to denote the transformation of features. For each image $V_i$ in the network, we sample a positive image $V_i^+$ to compose the positive pair, and we obtain its features from the final layer as $f_{t}(V_i)$. The similarity of the two images $V_i$ and $V_i^+$ is defined as follows:
\begin{equation} \label{equation_1}
\begin{aligned}
& Sim(V_i,V_i^+;\theta_{t})=
\frac{f_{t}(V_i;\theta_{t}) \cdot f_{t}(V_i^+;\theta_{t})}{\|f_{t}(V_i;\theta_{t})\|+\|f_{t}(V_i^+;\theta_{t})\|}
\end{aligned}
\end{equation}
where $\theta_{t}$ is the parameters shared in the Siamese-Triplet network. Similarly, given the image $V_i$, we can sample a negative sample $V_i^-$, and the similarity  $Sim(V_i,V_i^-;\theta_{t})$ can be calculated like Eq. (\ref{equation_1}).

To encode the network structure information into feature representation $f_{t}(\cdot)$, it should be enforced that the similarity of a given image to the positive image is larger than to the negative image, i.e., $Sim(V_i,V_i^+)>Sim(V_i,V_i^-)$. The loss of the ranking is formulated by the hinge rank loss as follows:
\begin{equation}
\begin{aligned}
\mathcal{L}_h(V_i,V_i^+,V_i^-;\theta_{t})= max[0,M
-Sim(V_i,V_i^+;\theta_{t})+Sim(V_i,V_i^-;\theta_{t})]
\end{aligned}
\end{equation}
where $M$ denotes the gap parameter between two similarities. We empirically set $M = 0.3$ in the experiments. Then our objective function for training is formulated as follows:
\begin{equation} \label{hinge_loss}
\begin{aligned}
& \mathcal{L}_h(V;\theta_{t})=\sum_{i=1}^N max[0,M-
Sim(V_i,V_i^+;\theta_{t})+Sim(V_i,V_i^-;\theta_{t})]
\end{aligned}
\end{equation}
where $N$ denotes the total number of the nodes. The $L_2$ normalization are replaced by dropout layers.

It is non-trivial to select the negative samples for learning to rank. We use the mini-batch Stochastic Gradient Descent (SGD) method to train the model. For each pair of $V_i$ and $V_i^+$, we randomly sample $K$ negative matches in the same batch \textit{B}, which obtains $K$ triplets of samples. For each triplet of samples, the gradients over the three samples are computed respectively and the parameters are updated using the back propagation method. To ensure that the pair of patches $V_i$ and $V_i^+$ can look up different negative matches each time, all the images are shuffled randomly after each iteration of training. For the experiments, we set $K=3$.

\subsection{Visual-Textual Attention Model} \label{sub:attention}

Attention is a mechanism which allows for an alignment of the input and output sequence \cite{2015Show-Xu}, by which the salient features are allowed to dynamically come to the forefront as needed. Recently, it has been proven to be beneficial for many vision related tasks, such as image captioning \cite{2015Show-Xu} and image question answering \cite{2016Stacked-Yang}. Different from these works, our attention model is formulated in the Siamese-Triplet architecture for multi-modal data. We use the attention model to capture the alignment between different data modalities, which can utilize the network information to semantically learn the mapping between words and image regions based on the deep model of Siamese-Triplet network.



\begin{figure}[htb]
  \centering
  \begin{minipage}[b]{0.35\textwidth}
  \includegraphics[width=0.95\textwidth]{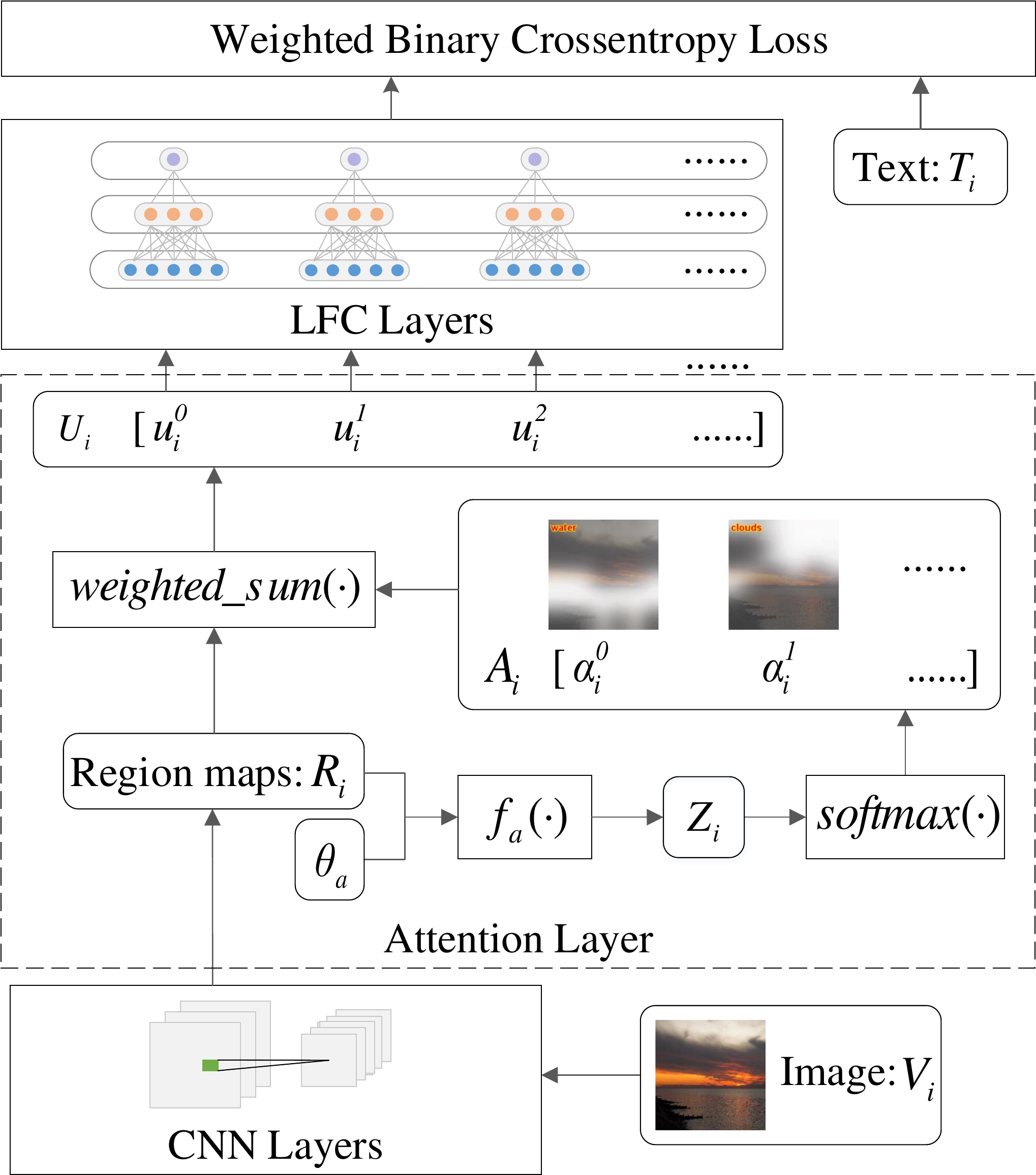}
  \end{minipage}
  \caption{The architecture of the visual-textual attention model. A rectangle block stands for a process or a function, while the rounded box stands for inputs, parameters or tensors.}
  \label{fig:attention_network}
\end{figure}

Given an image-text pair, our goal is to automatically find the relation between the words and the image regions. Let $ T_i =\{t^0_{i}, t^1_{i}, ..., $ $ t^k_{i}, ..., t^{L}_{i}\}, T_i\in \mathbb{R}^{L} $ denotes the text features of the $i th$ pair, which is a one-hot words vector expression with length $L$, where $k$ denotes for the index of the word. Let $V_i$ denotes the raw image corresponded to $T_i$. We use the deep Convolution Neural Networks (CNN) to obtain the image region maps $R_i=\{r_{i,0},r_{i,1},...,r_{i,j},...,$ $r_{i,D}\} \in \mathbb{R}^{D \times M}$ for $V_i$ as follows:

\begin{equation}
\begin{aligned}
& R_i = f_{c}(V_i;\theta_{c}),\ R_i \in \mathbb{R}^{D \times M}
\end{aligned}
\end{equation}
where $\theta_{c}$ is the parameters of the CNN layers, $j$ denotes the index of the region, $D$ is the dimension of image region, and $M$ is the dimension of the map.

In the attention model, a value between 0 and 1 is assigned to each image region $r_{i,j}$ based on its relevance to the word $t^k_{i}$. Formally, we aim to automatically generate the image attention values for the words as follows:
\begin{equation} \label{equation_6}
\begin{aligned}
& Z_i = f_{a}(R_i;\theta_{a}),\ Z_i \in \mathbb{R}^{L \times D}
\end{aligned}
\end{equation}
where $Z_i$ denotes the unnormalized word attention values for the region maps $R_i$ of the $i th$ pair. Following \cite{2015Show-Xu,2017Learning-Zhu}, $Z_i$ is spatially normalized with the softmax function to obtain the final word attention maps $A_i$,
\begin{equation}
\begin{aligned}
& \alpha^k_{i,j} = \frac{exp(z^k_{i,j})}{\sum_{j}{exp(z^k_{i,j})}},\ A_i \in \mathbb{R}^{L \times D}
\end{aligned}
\end{equation}
where $z^k_{i,j}$ and $\alpha^k_{i,j}$ represent the unormalized and normalized attention values at region $j$ in image $i$ for word $k$ respectively. If word $k$ is assigned to the input image, higher attention values should be given to the image regions related to it. The attention estimator $f_{a}(\cdot)$ can be calculated in many ways such as CNN in \cite{2017Learning-Zhu}. In this paper, it is modelled as a sequentially distributed full connected layer and computed as follows:
\begin{equation}
\begin{aligned}
& Z_i = tanh(wR^T_i+b),\ Z_i \in \mathbb{R}^{L \times D}
\end{aligned}
\end{equation}
where $w \in \mathbb{R}^{L \times M}$ and $b \in \mathbb{R}^{L}$ that compose the parameter set $\theta_{a}$ of the attention model in Eq. (\ref{equation_6}) and will be updated through back propagation. The tanh activation are used to make the model nonlinear.

Let $r_{i,j} \in \mathbb{R}^{M}$ denotes the visual feature vector of region $j$ in $R_i$. The normalized attention value is used as the weight to sum the features of the regions for each word $k$ to obtain the output features as follows:
\begin{equation}
\begin{aligned}
& u^k_{i} = \sum_j{\alpha^k_{i,j}r_{i,j}},\ U_i \in \mathbb{R}^{L \times M}
\end{aligned}
\end{equation}

The architecture of the attention model is illustrated in Figure \ref{fig:attention_network}. The equation above behaves somewhat like a weighted average pooling layer for each word. Compared with the original independent visual features shared by all words, the weighted visual feature mapping $u^k_{i}$ is more effective to reflect the image regions related to word $k$. The dimensionality of $U_i$ is $\mathbb{R}^{L \times M}$, while $T_i \in \mathbb{R}^{L}$. In order to make comparison between the visual output and text features, we stack several Locally Fully Connected (LFC) layers to obtain a $L$-dimensionality output of visual features. The LFC layers are locally fully connected for each word, and the parameter sets corresponding to different words are independent. That is, $u^k_{i}$ is only related to the correspondingly attended $t_i$ for each word $k$. Note that the last LFC layer has just one neuron for each word in the vocabulary, which sets the dimensionality of the final output equal to $L$. The activation sigmoid is used in the last LFC layer to normalise the feature representation for estimating words' confidence by comparing with the ground truth text vector. Let $Y_i \in \mathbb{R}^{L}$ be the final output of LFC layers:
\begin{equation}
\begin{aligned}
& Y_i = f_{l}(U_i;\theta_{l}),\ Y_i \in \mathbb{R}^{L}
\end{aligned}
\end{equation}

We make a pipeline of the three functions mentioned above to obtain a whole process from image input to $Y_i$ as:
\begin{equation}
\begin{aligned}
& Y_i = f_{w}(V_i;\theta_{w}),\ Y_i \in \mathbb{R}^{L}
\end{aligned}
\end{equation}
where $f_{w}$ is the pipeline of $f_{c}$, $f_{a}$ and $f_{l}$, and $\theta_{w}$ is the set of $\theta_{c}$, $\theta_{a}$ and $\theta_{l}$.

The parameters $\theta_{w}$ are learned by minimizing the weighted binary crossentropy loss between $Y_i$ and $T_i$,
\begin{equation} \label{crossentropy_loss}
\begin{aligned}
& \mathcal{L}_c(V,T;\theta_{w})
=\sum_{i=1}^N -(\lambda T_i \cdot log(Y_i))+(1-T_i) \cdot log(1-T_i) \\
=&\sum_{i=1}^N -(\lambda T_i \cdot log(f_{w}(V_i;\theta_{w})))
 +(1-T_i) \cdot log(1-f_{w}(V_i;\theta_{w}))
\end{aligned}
\end{equation}
where $N$ is the whole number of the pairs and $\lambda$ is a balance parameter. Since the number of zero elements in $T_i$ is much more than non-zero ones, it is reasonable to punish the false negatives more.

\subsection{A Joint Deep Embedding Model} \label{sub:joint}

The Siamese-Triplet neural network learns the embedding by exploiting the network structure information, and the visual-textural attention model exploits the fine-granularity relation between data modalities for embedding learning. Intuitively, we propose a joint deep embedding model to combine the two components, which simultaneously optimizes them. Specifically, we change the FC layers in the base network of Siamese-Triplet model to an attention layer and several LFC layers. Then, we formulate the loss function as the summation of the hinge rank loss Eq. (\ref{hinge_loss}) and the weighted binary crossentropy loss Eq. (\ref{crossentropy_loss}) as follows:
\begin{equation} \label{loss}
\begin{aligned}
& \mathcal{L}(V,T;\theta_{w})= \mathcal{L}_h(V;\theta_{w})
+ \beta \mathcal{L}_c(V,T;\theta_{w})
\end{aligned}
\end{equation}
where $\beta$ is a weight parameter. Since the Triplet network are added with the attention in the joint model, the parameter $\theta_{t}$ are replaced by $\theta_{w}$ which is shared in the whole model.

The computational complexity of the joint deep model is reduced greatly by using the positive sampling and negative smapling method. Assume the whole number of image-text pairs is $N$, the method in \cite{2015Heterogeneous-Chang} learns the network representation by iterating the whole network. It results in a computational complexity of $O(k(N \times N))$, where $k$ is the number of iterations. Our methods only sample several nodes for each node for parameters updating in each iteration. Thus the computational complexity is reduced to $O(k(N))$.




\section{EXPERIMENTS}

In this section, we conduct a set of experiments to analyse the effectiveness of the embedding model DMAN by evaluating its performance in two tasks, i.e., multi-label image classification and cross-modal search.

\subsection{Experimental Settings}

The experiments are conducted on three popular datasets collected from flickr, which have the groundtruth labels provided by human annotators. Based on the study of \cite{2012Image-McAuley} on these collections, we crawl the original images from the Flickr website. The details of these image collections are described below:
\begin{itemize}
\item[-] The NUS Web Image Database (NUS-WIDE) dataset \cite{2009NUS-Chua} is a web image dataset which contains 269,648 images. Among these images, 226,912 are available in the Flickr sources.
\item[-] The MIR Flickr Retrieval Evaluation (MIR) dataset contains one million images, but only 25,000 of them have been annotated \cite{2008MIR-Huiskes}. 13,368 of the annotated images are available in the Flickr sources.
\item[-] The PASCAL Visual Object Classes Challenge (PASCAL) datasets \cite{2010Pascal-Everingham} contains 9963 images. 9,474 of the annotated images are available in Flickr sources.
\end{itemize}

We preprocess these datasets as \cite{2015Heterogeneous-Chang}. First, since there are many noisy images that do not belong to any of their groundtruth labels, we remove these samples. After that, we use the most frequent 1,000 tags as our text vocaburary and construct a 1000-D 0-1 vector for the text content. We further remove those image-text pairs that do not contain any words in the vocaburary. Finally, we randomly sample the image-text pairs for training and testing with the ratio of 4:1. We construct a network by treating each image-text pair as a node, and establish an edge between two nodes once they share at least one label. For each node, at most 50 links are random sampled to construct the sparse adjacency matrix. We evaluate our framework in an out-of-sample strategy. The final statistics of these datasets are shown in Table  \ref{tab:experiment_datasets}. Note that 90000 nodes of NUS-WIDE are ramdomly sampled with 53,844 for training and 36,352 for testing as \cite{2015Heterogeneous-Chang} for fair comparison.

\begin{table}[htbp]
  \centering
  \caption{Datasets statistics.}
    \begin{tabular}{lrrr}
    \toprule
          & NUS   & MIR   & PASCAL \\
    \midrule
    \#image & 90,000 & 5,040  & 6,509 \\
    \#tag & 1,000  & 1,000  & 1,000 \\
    \#tag per image & 6.36  & 4.09  & 3.94 \\
    \#label & 81    & 14    & 20 \\
    \#label per image & 2.46  & 1.81  & 1.95 \\
    \midrule
    \#node (training) & 53,844 & 4,032  & 5,207 \\
    \#edge (training) & 2,682,005 & 216,318 & 284,961 \\
    \midrule
    \#node (testing) & 36,156 & 1,008  & 1,302 \\
    \bottomrule
    \end{tabular}%
  \label{tab:experiment_datasets}%
\end{table}%

In the experiments, we take the image with size $224\times224$ and channel RGB as visual input, and CNNs is used for visual feature extraction. Specifically, our CNN layers employ the VGG16 network \cite{2014Very-Simonyana} pretrained on ImageNet 2012 classification challenge dataset \cite{2009ImageNet-Deng} with Keras deep learning framework. Then we use the pool5's outputs as the visual features for image region maps, with size $49\times512$. We stack three LFC layers to the attention layer, with the dimensions of $1000 \times 128$, $1000 \times 32$ and $1000 \times 1$ respectively. As for the hyper-parameters $\lambda$ and $\beta$, we set the values with 10 and 1 that can obtain a relatively  good performance. In the training procedure, SGD is set with learning rate $0.01$, momentum 0.9 and nesterov=True. All of the implementation are trained on 2$\times$ NVIDIA GTX 1080. All the source codes of our models will be released upon the publication of this work.

\begin{table*}[htbp]
  \centering
  \caption{Multi-label classification results}
    \begin{tabular}{|c|lrrrrrrr|}
    \toprule
    Dataset & \multicolumn{1}{c}{Model} & \multicolumn{1}{c}{Micro-P} & \multicolumn{1}{c}{Micro-R} & \multicolumn{1}{c}{Micro-F1} & \multicolumn{1}{c}{Macro-P} & \multicolumn{1}{c}{Macro-R} & \multicolumn{1}{c}{Macro-F1} & \multicolumn{1}{c|}{mAP} \\
    \midrule
    \multirow{9}[2]{*}{NUS-WIDE} & CCA \cite{2004Canonical-Hardoon}   & -     & -     & -     & -     & -     & -     & 52.54\% \\
          & DT \cite{2012Transfer-Qi}    & -     & -     & -     & -     & -     & -     & 53.22\% \\
          & LHNE \cite{2015Heterogeneous-Chang}  & -     & -     & -     & -     & -     & -     & 53.32\% \\
          & HNE \cite{2015Heterogeneous-Chang}   & -     & -     & -     & -     & -     & -     & 54.99\% \\
          & KCCA \cite{2004Canonical-Hardoon}  & 75.65\% & 52.96\% & 62.30\% & 59.12\% & 46.58\% & 52.11\% & 52.91\% \\
          & DCCA \cite{2015Deep-Yan}  & 75.79\% & 54.48\% & 63.39\% & 61.21\% & 50.16\% & 55.14\% & 54.36\% \\
          & DMAN$_{Triplet}$ & 76.49\% & 54.30\% & 63.51\% & 59.69\% & 48.88\% & 53.75\% & 53.35\% \\
          & DMAN$_{Triplet+Text}$ & 77.00\% & 55.15\% & 64.26\% & 59.62\% & 51.40\% & 55.20\% & 54.84\% \\
          & DMAN & \textbf{77.25\%} & \textbf{56.37\%} & \textbf{65.18\%} & \textbf{62.64\%} & \textbf{52.62\%} & \textbf{57.19\%} & \textbf{57.52\%} \\
    \midrule
    \multirow{5}[2]{*}{MIR} & KCCA \cite{2004Canonical-Hardoon}  & 73.69\% & 72.55\% & 73.12\% & 72.41\% & 70.86\% & 71.63\% & 69.53\% \\
          & DCCA \cite{2015Deep-Yan}  & 77.21\% & 76.65\% & 76.93\% & \textbf{75.37\%} & 70.11\% & 72.64\% & 73.13\% \\
          & DMAN$_{Triplet}$ & 76.51\% & 71.75\% & 74.58\% & 73.58\% & \textbf{72.67\%} & 72.74\% & 72.05\% \\
          & DMAN$_{Triplet+Text}$ & 78.14\% & \textbf{77.66\%} & 77.90\% & 73.14\% & 71.08\% & 72.09\% & 74.40\% \\
          & DMAN & \textbf{79.14\%} & 77.45\% & \textbf{78.28\%} & 75.12\% & 72.11\% & \textbf{73.65\%} & \textbf{75.19\%} \\
    \midrule
    \multirow{5}[2]{*}{PASCAL} & KCCA \cite{2004Canonical-Hardoon}  & 78.55\% & 31.89\% & 45.36\% & 49.25\% & 35.77\% & 41.44\% & 43.59\% \\
          & DCCA \cite{2015Deep-Yan}  & 80.09\% & 35.27\% & 48.97\% & 53.91\% & 43.86\% & 48.37\% & 47.25\% \\
          & DMAN$_{Triplet}$ & 83.87\% & 30.32\% & 44.54\% & 56.16\% & 38.84\% & 45.92\% & 47.41\% \\
          & DMAN$_{Triplet+Text}$ & 79.19\% & 38.30\% & 51.63\% & 55.75\% & 41.12\% & 47.33\% & 49.54\% \\
          & DMAN & \textbf{86.88\%} & \textbf{44.14\%} & \textbf{58.53\%} & \textbf{62.23\%} & \textbf{46.85\%} & \textbf{53.45\%} & \textbf{54.76\%} \\
    \bottomrule
    \end{tabular}%
  \label{tab:multi_label_classification}%
\end{table*}%

\subsection{Baselines}

We evaluate the performance of DMAN by comparing it with state-of-the-art approaches introduced below:
\begin{itemize}
\item \textbf{CCA} \cite{2004Canonical-Hardoon}: The Canonical Correlation Analysis embeds two types of input data into a common latent space by optimizing an objective function with respect to their correlations.
\item \textbf{DT} \cite{2012Transfer-Qi}: A transfer learning method is proposed to bridge the semantic distances between image and text using latent embeddings.
\item \textbf{LHNE} \cite{2015Heterogeneous-Chang}: The linear version of Heterogeneous Network Embedding (HNE) \cite{2015Heterogeneous-Chang}.
\item \textbf{HNE} \cite{2015Heterogeneous-Chang}: Heterogeneous Network Embedding via Deep Architectures.
\item \textbf{KCCA} \cite{2004Canonical-Hardoon}: The kernel version of the Canonical Correlation Analysis.
\item \textbf{DCCA} \cite{2015Deep-Yan}: An image-text matching
approach based on deep canonical correlation analysis.
\item \textbf{DMAN$_{Triplet}$}: Only the image is used to construct the triplet neural network and the representation is learned from the network directly. It is used to evaluate the effectiveness of the triplet network model for embedding
\item \textbf{DMAN$_{Triplet+Text}$}: The text content is added by DMAN$_{Triplet}$, where the text content is combined using a full connected network rather than the attention network.
\end{itemize}

Among, the first 4 methods are introduced in \cite{2015Heterogeneous-Chang} and we will compare them with our model on the dataset NUS-WIDE for the two tasks as in their papers.


\begin{table*}[htbp]
  \centering
  \caption{Cross modal retrieval results}
    \begin{tabular}{|c|lrrrrr|}
    \toprule
    Dataset & \multicolumn{1}{c}{Model} & \multicolumn{1}{c}{Rank 1} & \multicolumn{1}{c}{Rank 5} & \multicolumn{1}{c}{Rank 10} & \multicolumn{1}{c}{Rank 20} & \multicolumn{1}{c|}{Rank 50} \\
    \midrule
    \multirow{8}[2]{*}{NUS-WIDE} & CCA \cite{2004Canonical-Hardoon}   & 21.05\% & 16.84\% & 18.95\% & 18.68\% & - \\
          & DT \cite{2012Transfer-Qi}    & 20.53\% & 25.26\% & 22.63\% & 22.37\% & - \\
          & LHNE \cite{2015Heterogeneous-Chang}  & 26.32\% & 21.05\% & 21.02\% & 22.27\% & - \\
          & HNE \cite{2015Heterogeneous-Chang}   & 36.84\% & 29.47\% & 27.89\% & 26.32\% & - \\
          & KCCA \cite{2004Canonical-Hardoon}  & 26.30\% & 27.69\% & 22.57\% & 20.36\% & 18.35\% \\
          & DCCA \cite{2015Deep-Yan}  & 35.99\% & 31.12\% & 29.15\% & 24.48\% & 25.32\% \\
          & DMAN$_{Triplet+Text}$ & 36.16\% & 28.57\% & 28.57\% & 27.92\% & 23.11\% \\
          & DMAN & \textbf{38.96\%} & \textbf{40.77\%} & \textbf{40.91\%} & \textbf{38.96\%} & \textbf{37.14\%} \\
    \midrule
    \multirow{4}[2]{*}{MIR} & KCCA \cite{2004Canonical-Hardoon}  & 50.83\% & 43.53\% & 37.60\% & 35.41\% & 30.50\% \\
          & DCCA \cite{2015Deep-Yan}  & 59.67\% & 54.12\% & 51.81\% & 41.89\% & 38.65\% \\
          & DMAN$_{Triplet+Text}$ & 64.50\% & 53.47\% & \textbf{54.16\%} & 52.18\% & 40.33\% \\
          & DMAN & \textbf{66.67\%} & \textbf{55.00\%} & 53.50\% & \textbf{52.18\%} & \textbf{43.67\%} \\
    \midrule
    \multirow{4}[2]{*}{PASCAL} & KCCA \cite{2004Canonical-Hardoon}  & 27.35\% & 28.06\% & 30.15\% & 25.12\% & 26.33\% \\
          & DCCA \cite{2015Deep-Yan}  & 42.12\% & 45.14\% & 44.96\% & 42.87\% & 41.89\% \\
          & DMAN$_{Triplet+Text}$ & 41.18\% & 40.00\% & 41.76\% & 43.20\% & 38.35\% \\
          & DMAN & \textbf{47.05\%} & \textbf{55.29\%} & \textbf{57.64\%} & \textbf{56.17\%} & \textbf{46.23\%} \\
    \bottomrule
    \end{tabular}%
  \label{tab:cross_search}%
\end{table*}%

\subsection{Multi-Label Classification}

All the datasets are multi-labeled with unbalanced distribution over the classes. A comprehensive introduction of evaluation metrics for multi-label classification is presented in \cite{2016Unified-Wu}. We employ macro/micro precision, macro/micro recall, macro/micro F1-measure, and Mean Average Precision (mAP) for performance evaluation. If the predicted label confidence for any label is greater than 0.5, the label is considered as positive. To ensure fair comparison, we use the neural network with 3 FC layers to learn a common classifier. After accomplishing the training process, we use the trained model to acquire the embeddings of the test set. Then we use the FC classifier mentioned above to train and test on each dataset.

\begin{figure*}[htb]
  \centering
  \begin{minipage}[b]{0.8\textwidth}
  \includegraphics[width=1\textwidth]{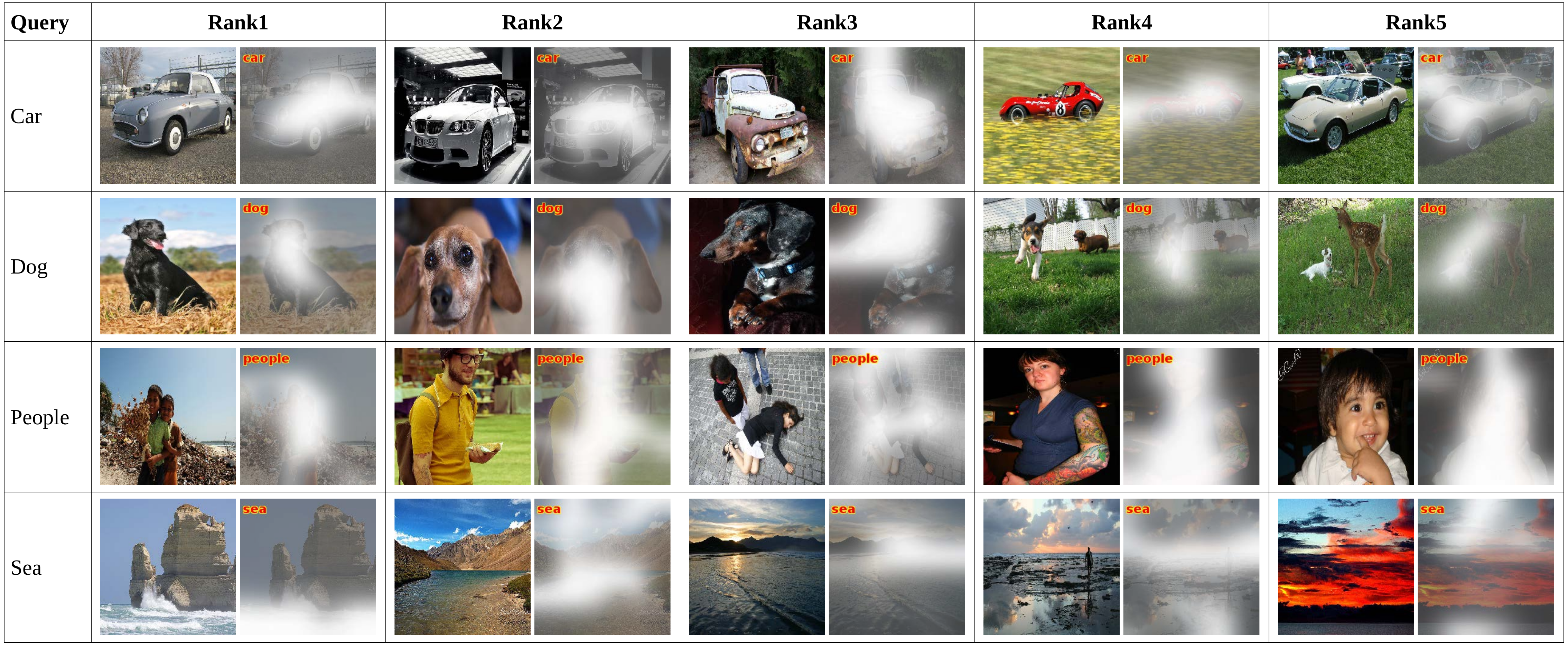}
  \end{minipage}
  \caption{Some results of cross-modal search and the visualization of learned attentions}
  \label{fig:experiment_attention}
\end{figure*}

Experimental results on the 3 datasets are shown in Table \ref{tab:multi_label_classification}. It shows that DMAN outperforms all of state-of-the-art models. First, from the results of NUS-WIDE, it can be concluded that the performance of DMAN$_{Triplet}$ is better than that of CCA, DT and LHNE on the metric of mAP, which validate the effectiveness of embedding using the triplet network model. By combining the text content, DMAN$_{Triplet+Text}$  almost reaches the score of HNE on the metric of mAP and transcend DMAN$_{Triplet}$ on all metrics, which confirms the importance of combining the multi-modal content for embedding learning. With the attention model, DMAN boosts mAP to 57.52\% compared with 54.99\% of HNE and makes an improvement on every metric comparing to DMAN$_{Triplet+Text}$. This is because that the attention model makes alignments between the multi-modal contents, which is useful to learn a more efficient representation of the multi-modal data. On the other side, HNE learns the features for image and text document independently, which is less effective to capture the correlation between different data modalities. Meanwhile, the other baselines also can not effectively exploiting the link information and the fine-granularity relations between different data modalities. The quantity of tags in PASCAL is less than those in other datasets, but the quality is better. Therefore, the improvement of DMAN$_{Triplet+Text}$ and DMAN on PASCAL is greater than it on other datasets.

\subsection{Cross-Modal Search}

To further demonstrate the superiority of DMAN, we compare it with the baselines in the task of cross-modal search as \cite{2015Heterogeneous-Chang}. In the datasets NUS-WIDE, MIR and PASCAL, there are 77/81, 12/14 and 17/20 of the groundtruth label words appearing in the text vector respectively. We manually construct 77, 12, and 17 query vectors with the dimensionality of 1000 for the three datasets respectively, by setting the corresponding label entry to one and the remaining entries to zero. Using the learned embedding function, we project the query vector to the latent space to retrieve all the image samples in the test set using the standard Euclidean distance. The average precision at rank \textit{k} (\textit{p}@\textit{k}) over all queries is reported in Table \ref{tab:cross_search}. On the dataset of NUS-WIDE, DMAN achieves about 10\% higher AP compared to HNE, and far surpass to CCA, DT, and LHNE. On all the three datasets, DMAN outperforms KCCA and DCCA greatly. It demonstrates the effectiveness of our model for cross-modal search. Meanwhile, the link information is also helpful to find the most similar images, which affects the performance of the methods that ignore the link information. For all of the datasets, we can find a substantial improvement of DMAN compared to DMAN$_{Triplet+Text}$, which provides evidence that the attention model is effective for cross-modal search.

Figure \ref{fig:experiment_attention} gives some samples of search results of  MIR. For each query, we present the top-5 ranked images and their corresponding regions aligned by the attention model. For the query ``Sea'', the aligned images have mismatched attention on white clouds and blue sky, which is because that the tags of ``sea'', ``cloud'' and ``sky'' frequently co-occur in the same images. For the other queries, our model draws the right attentions on the images and hence improves the performance.

\section{Conclusions}

In this paper, we explore learning social image embedding to capture both multimodal contents and network information for social media data applications, such as mutli-label classification and cross-modal search. A Deep Multimodal Attention Networks (DMAN) embedding model is proposed, in which a Siamese-Triplet neural network is designed to embed network information and the Visual-Textual Attention Model is proposed to capture the correlation between different data modalities. Then, a deep model is proposed to combing network information and content for embedding learning, in which the loss function is designed to simultaneously optimizing the Siamese-Triplet neural network and Visual-Textual Attention Model. The experiment results in multilabel classification and cross-modal search applications indicate that capturing the network structure and the correlation between different data modalities are helpful for social image embedding. In all of the experiments, our approach outperforms  state-of-the-art baselines in various evaluations.

This work is an effort to combine network structure for multimodal data embedding. It is different from current networking embedding researches that mainly explore the link information and can not effectively exploit the multi-modal content within each node. It is also different from the current image embedding methods that ignore the link information and are mainly task dependent, such as image caption and classification. In the future, we will explore to better model the social link information and design a more reasonable deep model to make the learned embedding more effective. Furthermore, we can incorporate the strength of relationship between images to better model the network.

\section{Acknowledgement}

This work was supported by the National Natural Science Foundation of China (Grand Nos. U1636211, 61672081, 61602237, 61370126), National High Technology Research and Development Program of China (No.2015AA016004) and Fund of the State Key Laboratory of Software Development Environment (No. SKLSDE-2017ZX-19).

\bibliographystyle{ACM-Reference-Format}

\end{document}